\documentclass[12pt]{iopart}

\usepackage{iopams}
\usepackage{amssymb}

\eqnobysec

\begin{document}

\title[Sylvester's theorem, Maxwell's multipoles and Majorana's sphere]{Canonical representation of spherical functions: Sylvester's theorem, Maxwell's multipoles and Majorana's sphere}

\author{M R Dennis}

\address{H H Wills Physics Laboratory, Tyndall Avenue, Bristol BS8 1TL, UK}

\begin{abstract}
Any eigenfunction of the laplacian on the sphere is given in terms of a unique set of directions: these are Maxwell's multipoles, their existence and uniqueness being known as Sylvester's theorem. Here, the theorem is proved by realising the multipoles are pairs of opposite vectors in Majorana's sphere representation of quantum spins. The proof involves the physicist's standard tools of quantum angular momentum algebra, integral kernels, and gaussian integration. Various other proofs are compared, including an alternative using the calculus of spacetime spinors.
\end{abstract}

\section{Introduction}

When manipulating functions on the sphere, it is convenient to have a functional representation whose behaviour under spatial rotations is transparent. 
This is not provided by the spherical harmonics $Y_j^m(\theta,\phi),$ which are required to transform as basis functions for the appropriate matrix representation of the rotation group SO(3). 
In his {\em Treatise on Electricity and Magnetism} \cite{maxwell:treatise1}, chapter IX, Maxwell found an approach to real sums of spherical harmonics, representing such functions as directional derivatives (``multipoles'') of $1/r;$ the resulting homogeneous polynomial in $x,y,z$ is the function represented by spherical harmonics on replacing the cartesian coordinates with the appropriate polar expression.
The directions of these derivatives rotate directly as vector-like objects.
Sylvester \cite{sylvester:note} proved that Maxwell's multipole representation is unique, this result (amongst others) being known as {\em Sylvester's theorem}.
The theorem has been discussed and proved by several authors: Courant and Hilbert \cite{ch:methods1} using Bezout's theorem from algebraic geometry; Backus \cite{backus:geometrical} using the first isomorphism theorem for rings; and Zheng and Zou \cite{zz:maxwell} using traceless symmetric tensors.

In this paper, I aim to provide an alternative, concrete proof of this result, using mathematical tools in every physicist's kit \footnote{It is unlikely that Sylvester would have approved, given his disparaging comments about Maxwell's method at the end of his note.}: quantum angular momentum algebra, integral kernels, and gaussian integration.
In section \ref{sec:comparison}, comparison is made between these various proofs. 
The main object utilised is Majorana's sphere representation for quantum spins \cite{majorana:atomi}.
In this representation, a spin state (of definite spin $j$) is represented by $2j$ points (unit vectors) on the Riemann sphere, generalising the familiar Bloch sphere representation for spin $1/2.$
Majorana himself introduced his sphere representation to aid calculation of transition probabilities in quantum mechanics \cite{majorana:atomi}.
It has been popularised more recently by Penrose \cite{penrose:emperors}, \cite{penrose:shadows} (particularly appendix C), and has been used to describe some Bell-type inequalities \cite{zp:bell}, to quantify chaos and calculate geometric phases for high quantum spins \cite{hannay:zero, hannay:berry}, and characterise polarization in 3-dimensional electromagnetic and other fields \cite{hannay:majorana, dennis:thesis}.
The main objective of the present work is to demonstrate that in the special case of a real state of fixed spin, the Majorana vectors {\em are} the Maxwell multipole directions.
Majorana's representation is more general, applying to both integer and noninteger spins and complex functions, but does not, in general, have the concrete multipole interpretation described below.

Maxwell's multipole representation represents a real function $f(\theta,\phi),$ with definite angular momentum $j(j+1)$  - that is, it is eigenfunction of the laplacian on the unit sphere with eigenvalue $-j(j+1)$ - as the direction-dependent part of a multipole derivative; i.e. there are $j$ unit vectors $\bi{u}_1, \dots, \bi{u}_j$ such that 
\begin{equation}
   f = D_{\bi{u}_1} \dots D_{\bi{u}_j} \frac{1}{r},
   \label{eq:maxmulti}
\end{equation}
where $D_{\bi{u}_i}$ represents the directional derivative operator $\bi{u}_i \cdot \nabla,$ and $r = \sqrt{x^2 + y^2 + z^2}.$
It is, of course, straightforward to show that a function of this form is an eigenfunction of the laplacian (see, for instance, \cite{rose:elementary}, section 28); however, it requires more work to show that any eigenfunction can be put into this form - this is Sylvester's theorem.

A more useful form of (\ref{eq:maxmulti}) is obtained by finding the derivatives explicitly, then expressing geometrically in terms of vectors (e.g. via a homogeneous polynomial representation in cartesian coordinates $x,y,z$). 
In effect, the directional derivatives are replaced by scalar products with the corresponding vector, and there is an additional term consisting of $r^2$ times a complicated combination $F$ of the components of the $\bi{u}_i,$
\begin{equation}
   f(\theta, \phi) = C (\bi{u}(\theta,\phi) \cdot \bi{u}_1) \dots  (\bi{u}(\theta,\phi) \cdot \bi{u}_j) + r^2 F.
   \label{eq:maxmulti1}
\end{equation}
Here, the unit vector $\bi{u}(\theta,\phi)$ is the point on the sphere given by the spherical coordinates $\theta, \phi,$ and here and hereafter, $C$ is an unspecified constant factor.
The equivalence between the representations (\ref{eq:maxmulti}) and (\ref{eq:maxmulti1}) is demonstrated in \ref{sec:equiv} using Fourier integration methods.
The latter form of Sylvester's theorem will be proved here.

The Maxwell multipoles have simple forms for the real parts of the spherical harmonics $Y_j^m(\theta, \phi)$: for zonal harmonics ($m = 0$), the $j$ multipoles are aligned in the $z$-direction, whereas the tesseral harmonics ($m > 0$) have $m$ multipoles in the $xy$-plane arranged in a regular $2m$-gon, and the remaining $j-m$ aligned in the $z$-direction.
A further simple example is the case when $j = 2,$ where the function is equivalent to a traceless symmetric tensor.
On adding a sufficiently large, arbitrary constant times the identity matrix, $f$ can be represented geometrically by an ellipsoid: the two planes in which the projections of the ellipsoid are circular are perpendicular to the Maxwell multipoles in this case.

\section{Quantum notation, spherical harmonics and spin coherent states}

Notation from the quantum mechanics of spin will be used, although it applies for general functions on the sphere. 
This provides a convenient representation of the linear algebra of the underlying mathematical structures, i.e. the irreducible representations of the rotation group SO(3) and the special linear group SL(2,$\mathbb{C}$).
$|\psi \rangle$ denotes an abstract ket state with definite integer or half-integer spin $j,$ fixed once and for all (i.e. it is an eigenket of total angular momentum $\hat{J}^2$).
$j$ will be assumed to be an integer unless stated otherwise.

With respect to a certain coordinate frame, $| \psi \rangle$ may be written
\begin{equation}
   | \psi \rangle = \sum_{m=-j}^j \psi_m | m, j \rangle,
   \label{eq:psidecomp}
\end{equation}
where $| m, j \rangle$ is an eigenket of the angular momentum component operator $\hat{J}_z$ with eigenvalue $m,$ and where no confusion will ensue, $| m, j \rangle$ is written $| m \rangle,$ and the same applies for the corresponding bras.
The phase convention used is the standard Condon-Shortley convention given by applying successive raising operators to $| -j \rangle.$
These states will often be rotated by 3-dimensional rotation operators; the rotation operator taking the unit vector in the $+z$-direction, $\bi{u}(0),$ to the unit vector $\bi{u}(\theta,\phi),$ is denoted $\hat{R}_z(\phi) \hat{R}_y(\theta).$ 
In terms of Euler angles, the vector is first rotated through $\theta$ in the $y$-direction, then through $\phi$ in the $z$-direction. 
The notation will be abused, and rotation of the corresponding bra-ket states will be represented by the same operators:
\begin{equation}
   \hat{R}_z(\phi) \hat{R}_y(\theta) | m \rangle = | m; \theta, \phi \rangle
   \label{equation:ketrot}
\end{equation}
where the ket on the right hand side is an eigenket of the angular momentum component operator $\bi{u}(\theta,\phi)\cdot \hat{\bi{J}}$ with eigenvalue $m.$

It may be shown, either from group theory \cite{wigner:group}, or other methods \cite{fls:feynman3, sakurai:modern}, that for $j$ integer or half integer, the rotation operator matrix elements  (Wigner $\mathcal{D}$-functions) are given by
\begin{eqnarray}
\fl   \langle m | \hat{R}_z(\phi) \hat{R}_y(\theta) | m' \rangle & = \mathcal{D}^{(j)}_{m,m'}(\phi\theta0) \nonumber \\
   & = \exp(-\rmi m \phi) \sqrt{(j+m')!(j-m')!(j+m)!(j-m)!} \nonumber \\
   & \qquad \qquad \times \sum_k \frac{(-1)^{k-m'+m} c^{2j+m'-m-2k} s^{m-m'+2k}}{(m-m'+k)!(j+m'-k)!(j-m-k)! k!}
   \label{eq:wignerd}
\end{eqnarray}
where in the sum, $c = \cos\theta/2, s = \sin \theta/2$ and the index $k$ runs through all integers with nonnegative arguments in the denominator factorials.
Functions of spatial coordinates (i.e. in `coordinate representation') depend on these matrix elements, as we shall see.
Here and in the following, the active rotation convention of \cite{sakurai:modern, vmk:quantum} will be used, rather than the passive convention of \cite{wigner:group, fls:feynman3, ll:quantum}.

A standard result (see, for example, \cite{vmk:quantum} equation 4.17a, \cite{sakurai:modern} equation (3.6.52)), on putting $m' = 0$ in (\ref{eq:wignerd}) and conjugating, gives the `fully folded' representation of the ket $| m \rangle$ as a spherical harmonic,
\begin{equation}
   \langle 0; \theta, \phi | m \rangle = \mathcal{D}^{(j)}_{m,0}(\phi\theta0)^{\ast} = \sqrt{\frac{4\pi}{2j+1}} Y_j^m(\theta,\phi).
   \label{eq:shdef}
\end{equation}
The term `fully folded' refers to the fact that the $m' = 0$ state has been used in this representation; it is equivalent to mapping the spin states into basis functions for the rotation group SO(3).
This result gives an equivalence between general spin $j$ states and functions decomposed in terms of spherical harmonics:
\begin{eqnarray}
   \psi(\theta, \phi) & = \langle 0; \theta, \phi | \psi \rangle \nonumber \\
   & = \sqrt{\frac{4\pi}{2j+1}} \sum_{m = -j}^j \psi_m Y_j^m(\theta,\phi).
   \label{eq:psish}
\end{eqnarray}

Directions on the sphere, instead of being represented by the spherical coordinates $\theta, \phi,$ can be represented by the stereographic coordinate in the complex plane
\begin{equation}
   \zeta(\theta, \phi) \equiv \exp(\rmi \phi) \tan\theta/2,
   \label{eq:zetadef}
\end{equation}
which is a complex number (possibly $\infty$) obtained by stereographic projection from the south pole.
When the context is clear, this coordinate will be denoted simply by $\zeta.$
It will often be useful to use this stereographic representation; when appropriate, $|m; \theta, \phi \rangle$ shall be written $|m; \zeta \rangle,$ etc.
It is easy to find (\ref{eq:wignerd}) when $m' = -j$ (and $j$ is integer or half-integer), giving the {\em spin coherent state} \cite{ks:primer}:
\begin{eqnarray}
\fl   \langle -j; \zeta | m \rangle & = \exp(\rmi m \arg \zeta) \sqrt{(2j)!(j+m)!(j-m)!}  \frac{(-1)^{j+m} c^{j-m} s^{j+m}}{ (j+m)! (j-m)!} \nonumber \\
   & = \sqrt{\frac{(2j)!}{(j+m)!(j-m)!}} c^{2j} \exp(-\rmi j \arg \zeta) (-1)^{j+m} \left[ \exp(\rmi \arg \zeta) \zeta \right]^{j+m} \nonumber \\
   & = \frac{\exp(-\rmi j \arg\zeta)}{(1+ |\zeta|^2)^j} (-1)^{j+m} \left(\begin{array}{c} 2j \\ j+m \end{array} \right)^{1/2} \zeta^{j+m}.
   \label{eq:scs}
\end{eqnarray}
Thus, in the `fully extended representation' (choosing $m' = -j$), $| m \rangle$ is represented as an $m$-dependent monomial of $\zeta$ with positive power, times an $m$-independent function of $\zeta$ and a numerical factor.
For brevity, this numerical factor is written
\begin{equation}
   \mu_m \equiv (-1)^{j+m} \left(\begin{array}{c} 2j \\ j+m \end{array} \right)^{1/2}.
   \label{eq:majfactor}
\end{equation}
In the fully extended representation, states are written as coherent basis functions for the linear group SL(2,$\mathbb{C}$). 
A general ket is represented by its {\em Majorana function} \cite{majorana:atomi, penrose:shadows}
\begin{eqnarray}
   p_{\psi}(\zeta) & \equiv \langle -j; \zeta | \psi \rangle \nonumber \\
   & = \frac{\exp(-\rmi j \arg \zeta)}{(1+|\zeta|^2)^j} \sum_{m = -j}^j \psi_m \mu_m \zeta^{j+m}.
   \label{eq:majorana}
\end{eqnarray}
The function $p_{\psi}(\zeta)$ is a polynomial in $\zeta,$ times a prefactor which depends nonanalytically on $\zeta,$ and all $| \psi \rangle$-dependence is in the polynomial part.
The polynomial will be called the {\itshape Majorana polynomial}; the zeros of $p_{\psi}(\zeta)$ are the zeros of the polynomial.
If $m'$ were taken to be $+j$ rather than $-j$ in (\ref{eq:scs}), the polynomial in (\ref{eq:majorana}) would depend on $\zeta^{\ast},$ not $\zeta;$ this representation was used by Penrose \cite{penrose:shadows} with north-pole stereographic projection.
Unlike the fully folded representation (\ref{eq:psish}), the Majorana representation can be used when $j$ is an integer or a half integer.

The abstract ket $| \psi \rangle$ can therefore be represented by the fully folded state (\ref{eq:psish}), given in terms of spherical harmonics, or the fully extended state, whose details depend on the Majorana polynomial in the stereographic coordinate $\zeta.$
It is useful to transform between these two representations, and to do this, we introduce the {\em extending kernel} (where $\zeta$ is assumed independent of $\theta, \phi$),
\begin{eqnarray}
   K(\zeta | \theta, \phi) & \equiv \frac{2j+1}{4\pi} \langle -j; \zeta |0; \theta, \phi  \rangle \nonumber \\
   & = \frac{2j+1}{4\pi} \sum_{m=-j}^j \langle -j; \zeta | m \rangle \langle m | 0; \theta, \phi \rangle \nonumber \\
   & = \sqrt{\frac{2j+1}{4\pi}} \frac{\exp(-\rmi j \arg \zeta)}{(1+|\zeta|^2)^j} \sum_{m=-j}^j Y_j^{m\ast}(\theta,\phi) \mu_m \zeta^{j+m}.
   \label{eq:kerdef}
\end{eqnarray}
In the second line, a resolution of the identity has been used, and in the third, (\ref{eq:psish}) and (\ref{eq:majorana}).
The integral kernel can therefore be written as a Majorana function, whose coefficients are the conjugate 
spherical harmonics, depending on $\theta$ and $\phi.$
It is straightforward to verify that $K(\zeta|\theta,\phi)$ indeed transforms between the representations:
\begin{equation}
   \int_0^{2\pi} \rmd \phi \int_0^{\pi} \rmd \theta \, \sin \theta\,  \psi(\theta,\phi) K(\zeta |\theta, \phi) = p_{\psi}(\zeta).
   \label{eq:kerver}
\end{equation}
Its inverse, the folding kernel $K^{\dagger}(\theta,\phi | \zeta )$ is defined
\begin{equation}
   K^{\dagger}(\theta,\phi | \zeta ) = \frac{2j+1}{4\pi} \langle 0; \theta,\phi  | -j; \zeta \rangle.
   \label{eq:invker}
\end{equation}
This kernel provides the inverse of (\ref{eq:kerver}):
\begin{equation}
   \int_{\mathbb{C}} \frac{\rmd\zeta \rmd \zeta^{\ast}}{(1+|\zeta|^2)^2} p_{\psi}(\zeta) K^{\dagger}(\theta,\phi | \zeta ) = \psi(\theta,\phi).
   \label{eq:invkerver}
\end{equation}

\section{The Majorana sphere}\label{sec:majorana}

The equation for the Majorana polynomial (\ref{eq:majorana}) applies for any integer or half-integer $j.$ In particular, if $j = 1/2,$ the Majorana polynomial is linear in $\zeta.$
Its single root is $\psi_{-1/2}/\psi_{+1/2},$ which may be inverse stereographically projected onto the Riemann sphere. 
In quantum mechanics, this representation of a spin $1/2$ spinor by a point on the unit sphere is known as the Bloch sphere representation; it is useful because the point on the Bloch sphere is independent of the overall normalisation and phase of the state.

The Majorana sphere is the generalisation of the Bloch sphere construction to polynomials of arbitrary order $2j.$ 
By the fundamental theorem of algebra, the Majorana polynomial has $2j$ complex roots, which are inverse stereographically projected onto the Riemann sphere, giving $2j$ unit vectors, which are points on the {\em Majorana sphere}.
These points rotate rigidly when the state is rotated, and such rotations correspond to unitary M{\"o}bius transformations on $\zeta$ in the Majorana function. 
These unit vectors will be called {\em Majorana vectors.}

Therefore, 
\begin{eqnarray}
   p_{\psi}(\zeta) & = \frac{\exp(-\rmi j \arg \zeta)}{(1+|\zeta|^2)^j} \sum_{m=-j}^j \psi_m \mu_m \zeta^{j+m} \nonumber \\
   & = \frac{\exp(-\rmi j \arg\zeta)}{(1+|\zeta|^2)^j} (-1)^{2j} \psi_j \prod_{n=1}^{2j}(\zeta - \zeta_{n}),
   \label{eq:majfactorise}
\end{eqnarray}
where the $2j$ roots are labelled $\zeta_{n},$ with spherical coordinates $\theta_{n}, \phi_{n}.$ 
The terms in this product are all linear in $\zeta,$ and may therefore be viewed as spin $1/2$ matrix elements.
Including the appropriate normalisation, and writing in the angular momentum $j,$ this means that
\begin{equation}
\fl   \langle -j, j; \zeta  |\psi \rangle  = p_{\psi}(\zeta) 
   = \frac{(-1)^{2j} \psi_j}{\prod_{n=1}^{2j} (1 + |\zeta_n|^2)^{-1/2}} \prod_{n=1}^{2j} \left\langle \left. -\frac{1}{2}, \frac{1}{2} ; \zeta  \right| -\frac{1}{2}, \frac{1}{2} ; \zeta_{n}\right\rangle.
   \label{eq:majfactorise1}
\end{equation}
This may therefore be called Majorana's theorem. It states that the coherent state wavefunction of a given state of spin $j$ (integer or half integer) is proportional to the product of $2j$ spin $1/2$ coherent states.
The proportionality factor depends on $\psi_j$ (whose modulus may be expressed in terms of the symmetric polynomials of the Majorana polynomial, and their conjugates), and the roots $\zeta_n.$
It is invariant with respect to unitary M{\"o}bius transformations of the roots, thus the right hand side of (\ref{eq:majfactorise1}) rotates rigidly, according to the rotation of spin 1/2 states only.
This result does not appear to be particularly well-known in the literature of quantum angular momentum.

An important observation, made by \cite{backus:geometrical, zz:maxwell} (for polynomials analogous to the Majorana polynomial), is that the Majorana vectors form antipodal pairs for a state whose fully folded representation is real, as will now be shown.
Let $f(\theta, \phi)$ be a real function on the sphere, with definite $f,$ given in terms of spherical harmonics
\begin{equation}
   f(\theta,\phi) = \sqrt{\frac{4\pi}{2j+1}} \sum_{m=-j}^j a_m Y_j^m(\theta,\phi).
   \label{eq:fdef}
\end{equation}
Spherical harmonics are related to their conjugates by the following standard expression:
\begin{equation}
   Y_j^{m \ast}(\theta,\phi) = (-1)^m Y_j^{-m}(\theta,\phi).
   \label{eq:shconj}
\end{equation}
Reality of $f$ therefore requires that the complex coefficients $a_m$ satisfy an equivalent condition:
\begin{equation}
   a_{m}^{\ast} = (-1)^m a_{-m}.
   \label{eq:amconj}
\end{equation}
The Majorana function $p_f(\zeta)$ corresponding to $f$ is given by (\ref{eq:majorana}), with $\psi_m$ replaced by $a_m.$
The stereographic coordinate representing the antipodal point to $\zeta$ is $-1/\zeta^{\ast},$ as may be directly verified from (\ref{eq:zetadef}).
Putting this into the Majorana function,
\begin{eqnarray}
\fl   p_f(-1/\zeta^{\ast}) & = \frac{\exp(-\rmi j \arg \zeta) |\zeta|^{2j}}{(1+|\zeta|^2)^j} \sum_{m=-j}^j a_m \mu_m (-1)^{j+m} \zeta^{\ast -j-m} \nonumber \\
   & = \frac{\exp(-\rmi j \arg \zeta) (-1)^j}{(1+|\zeta|^2)^j} \sum_{m=-j}^j (-1)^m a_m \mu_m |\zeta|^{j-m} \exp(\rmi (j+m)\arg\zeta) \nonumber \\
   & = \frac{\exp(\rmi j \arg \zeta) (-1)^j}{(1+|\zeta|^2)^j} \sum_{m=-j}^j (-1)^m a_{-m}^{\ast} \mu_{-m} \zeta^{\ast j-m} \nonumber \\
   & = (-1)^j p_f(\zeta)^{\ast},
   \label{eq:repol}
\end{eqnarray}
where in the third line, the identity (\ref{eq:amconj}) and the fact that $\mu_{-m} = \mu_{m}$ have been used, and in the fourth, $-m$ has replaced $m$ in the summing index.
This result shows that $\zeta$ is a root of $p_f$ if and only if $-1/\zeta^{\ast}$ is; the Majorana vectors of real functions are antipodal.
Since the extending kernel $K(\zeta|\theta,\phi)$ can be written in Majorana form (\ref{eq:kerdef}), with coefficients $Y_j^m$ satisfying (\ref{eq:amconj}), this shows that the Majorana vectors of the kernel are arranged in antipodal pairs.

In quantum mechanics, such `real' states occur in the presence of time reversal symmetry, since it can be shown that the time reversal operator antipodises the Majorana vectors of the state \cite{dennis:thesis}.
Thus, in situations such as Kramers' degeneracy \cite{ll:quantum}, when $j$ is an integer, the states are real and there is no degeneracy, whereas if $j$ is a half integer, the states are twofold degenerate with mutually antipodal Majorana vectors.

\section{The Majorana polynomial for spin 1 and 3-dimensional vectors}

The simplest Majorana function which has a folded representation occurs in the case $j = 1,$ when the polynomial is quadratic.
In this case,
\begin{eqnarray}
   p_{\psi}(\zeta) & = \frac{\exp(-\rmi \arg\zeta)}{(1+|\zeta|^2)^j} \left(\psi_{-1} - \sqrt{2} \psi_0 \zeta + \psi_{+1} \zeta^2\right) \nonumber \\
   & = \frac{\exp(-\rmi \arg\zeta) \psi_{+1}}{(1+|\zeta|^2)^j} (\zeta - \zeta_-)(\zeta - \zeta_+)
   \label{eq:ppsi1}
\end{eqnarray}
where the roots $\zeta_{\pm}$ are
\begin{equation}
   \zeta_{\pm} = \frac{\psi_0 \pm \sqrt{\psi_0^2-2\psi_{+1} \psi_{-1}}}{\sqrt{2} \psi_{+1}}.
   \label{eq:ppsiroots}
\end{equation} 

This can be seen more clearly if the state represented by a spherical basis $|m\rangle$ is replaced by a (generally complex) vector $\bi{v}$ in a cartesian basis, according to the standard transformation
\begin{equation}
   \left(\begin{array}{c} v_x\\ v_y\\ v_z\end{array}\right) = \frac{1}{\sqrt{2}}\left(\begin{array}{ccc} -1 & 0 & 1 \\ -\rmi & 0 & -\rmi \\ 0 & \sqrt{2} & 0 \end{array}\right) \left(\begin{array}{c} \psi_{+1}\\ \psi_0 \\ \psi_{-1} \end{array}\right).
   \label{eq:s2c}
\end{equation}
From here, it is clear that $\bi{v}$ is real if and only if $\psi_{+1}^{\ast} = - \psi_{-1}$ and $\psi_0$ is real.
Since the matrix is unitary, $\bi{v}$ is normalised: $\bi{v}^{\ast} \cdot \bi{v} = 1.$
In a cartesian basis, the pair of roots (\ref{eq:ppsiroots}) becomes
\begin{equation}
   \zeta_{\pm} = \frac{v_z \pm \sqrt{\bi{v} \cdot \bi{v} }}{-v_x+\rmi v_y}.
   \label{eq:ppsirootc}
\end{equation}

If the state is real, with spherical coefficients $a_{+1}, a_0, a_{-1}$ satisfying (\ref{eq:amconj}) (so $\bi{v}$ is also real), the expression for pair of roots simplifies:
\begin{equation}
   \zeta_{\pm} = \frac{\sqrt{2} a_{-1}}{a_0 \mp 1} = \frac{v_x+\rmi v_y}{v_z \mp 1}.
   \label{eq:ppsirootsre}
\end{equation}
This is the familiar stereographic representation for the pair of antipodal unit vectors $\pm \bi{v}.$
This implies that any real state of spin 1 corresponds to a ket $| 0,1; \theta, \phi\rangle,$ where $\theta,\phi$ are the spherical coordinates of one of the pair of antipodal Majorana vectors; this is a real, 3-dimensional vector up to a sign.

It is therefore natural to ask what the ket $| +1, 1; \zeta(\theta, \phi) \rangle$ corresponds to in 3-dimensional geometry.
Such a state is a rotation of the basis state $| +1,1 \rangle,$ which, by (\ref{eq:scs}), has Majorana function
\begin{equation}
   \langle -1,1; \zeta | +1,1 \rangle = \frac{\exp(-\rmi \arg\zeta)}{(1+|\zeta|^2)} \zeta^2.
   \label{eq:maj2zero}
\end{equation}
This has a repeated root at zero, corresponding to a pair of Majorana vectors in the $+z$-direction (thus these spin representations can be used to define spin 1/2 spinors \cite{cartan:theory}).
Therefore, the two Majorana vectors for $| +1, 1; \zeta(\theta,\phi)\rangle$ are the same, and are in the direction $\bi{u}(\theta, \phi)$ since the Majorana sphere rotates rigidly.
The case of repeated roots corresponds to the discriminant of the Majorana polynomial being zero; by (\ref{eq:ppsiroots}) and (\ref{eq:ppsirootc}), this implies that, in a cartesian frame, $\bi{v}(\theta,\phi) \cdot \bi{v}(\theta,\phi) = 0$ (note there is no conjugation).
Such vectors are called {\em nilpotent} (or {\em isotropic}) \cite{cartan:theory, bd:324}; the nilpotent cartesian vector corresponding to $| +1,1; \zeta\rangle$ will be denoted $\bnu(\zeta).$
It can also be shown \cite{bd:324, dennis:thesis} that the real vector $\rmi \bnu(\zeta)^{\ast} \times \bnu(\zeta)$ is parallel to the $\zeta$-direction; such vectors in the electromagnetic field $\bi{E} + \rmi \bi{B}$ have relativistically distinguished status (e.g. \cite{bb:vortex,dennis:thesis} and \cite{pr:spinors2} p258).

Any inner product involving spin 1 states can therefore be written in terms of inner products of cartesian vectors (which may be real, nilpotent or general complex).
This implies that the vector corresponding to the ket $| -1,1; \zeta \rangle$ is $\bnu(\zeta)^{\ast},$ giving the required identity $\langle -1,1; \zeta | +1,1; \zeta \rangle = \bnu(\zeta) \cdot \bnu(\zeta) = 0.$

For any real state, corresponding to $| 0, 1; \theta,\phi \rangle,$ the normalisation $|a_{+1}|^2 + |a_0|^2+|a_{-1}|^2 = 1$ implies that, up to a phase factor 
\begin{equation}
   a_{+1} = \frac{\sqrt{3}}{2\sqrt{\pi}} Y_1^{1\ast}(\theta,\phi);
   \label{eq:spin1const}
\end{equation}
this constant appears as a multiplying factor in the Majorana function (\ref{eq:ppsi1}), and will be used in the next section.

\section{Analogue of Majorana's theorem for real states}

The form of the Majorana polynomial for general basis states is given by (\ref{eq:scs}): the polynomial for $| m \rangle$ has $j+m$ vectors in the $+z$ direction $\bi{u}(0),$ and $j-m$ in the $-z$-direction $\bi{u}(\pi).$ 
As described in the previous section, since the Majorana sphere rotates rigidly, the Majorana function $| m; \theta, \phi \rangle$ has $j+m$ Majorana vectors in the direction $\bi{u}(\theta,\phi),$ and $j-m$ vectors in the opposite direction $-\bi{u}(\theta, \phi) = \bi{u}(\pi - \theta,\phi + \pi).$
When $j$ is an integer, the extending kernel $K(\zeta|\theta, \phi)$ of (\ref{eq:kerdef}) exists, and by its definition as a Majorana function, it has $j$ repeated Majorana vectors in $\bi{u}(\theta,\phi)$ and $j$  in $-\bi{u}(\theta,\phi).$
Therefore
\begin{eqnarray}
\fl   \sqrt{\frac{4\pi}{2j+1}} K(\zeta | \theta, \phi) & = \frac{\exp(-\rmi j \arg\zeta)}{(1+|\zeta|^2)^j} \sum_{m=-j}^j Y_j^{m \ast}(\theta,\phi) \mu_m \zeta^{j+m} \nonumber \\
   & = \frac{\exp(-\rmi j \arg\zeta) Y_j^{j \ast}(\theta,\phi)}{(1+|\zeta|^2)^j} \left(\zeta - \zeta(\theta,\phi) \right)^j \left(\zeta + 1/\zeta^{\ast}(\theta, \phi) \right)^j \nonumber \\
   & = \frac{Y_j^{j \ast}(\theta,\phi)}{(\sqrt{3} Y_1^{1\ast}(\theta, \phi)/2\sqrt{\pi})^j} \langle -1, 1; \zeta | 0,1; \theta, \phi \rangle^j \nonumber \\
   & = C \left(\bi{u}(\theta,\phi) \cdot \bnu(\zeta)\right)^j.
   \label{eq:kerdecomp}
\end{eqnarray}
As before, $C$ denotes an unspecified numerical constant, and inner products have been rewritten in a cartesian basis.
In the penultimate line, the appropriate normalisation factor for the $j = 1$ polynomial from (\ref{eq:spin1const}) was used, and the $\theta,\phi$ dependence in the prefactor here cancels.
A consequence of the factorisation in the second line implies that the conjugated spherical harmonics $Y_j^{m \ast}$ may be constructed as $Y_j^{j \ast}$ times the appropriate symmetric polynomials in the roots $\zeta(\theta, \phi)$ and $-1/\zeta(\theta,\phi)^{\ast}$ and a numerical factor.
The final line of (\ref{eq:kerdecomp}) shows that any extending kernel for integer $j$ is proportional to the $j$th power of the dot product of a real vector and a nilpotent one.

The Majorana polynomial $p_f(\zeta)$ for the real function $f$ (\ref{eq:repol}), may be written in a similar way to (\ref{eq:kerdecomp}), since its Majorana vectors form antipodal pairs. 
Choosing and labelling a vector $\bi{u}_n = \bi{u}\bi(\theta_n,\phi_n)$ from each pair, with $n$ running from 1 to $j:$
\begin{eqnarray}
   p_f(\zeta) & = C \prod_{n = 1}^{j} \langle -1,1; \zeta | 0, 1; \theta_{n}, \phi_{n} \rangle \nonumber \\
   & = C' \prod_{n = 1}^{j} \bi{u}_n \cdot \bnu(\zeta).
   \label{eq:pfdecomp}
\end{eqnarray}
Changing the representative from a pair changes the sign of the unspecified numerical constants $C, C'.$
The first line of this equation is similar to Majorana's theorem (\ref{eq:majfactorise1}), although in the case of a real function, the decomposition is possible in terms of spin 1 coherent states, as well as spin 1/2.
The second line resembles Maxwell's multipole decomposition (\ref{eq:maxmulti1}), although it is the Majorana function $p_f(\zeta)$ that is given here, and it is written in terms of the nilpotent vector $\bnu(\zeta),$ not the real vector $\bi{u}(\theta, \phi);$ furthermore, there is no additional summand (this vanishes, since $\bnu\cdot\bnu = 0$).
This tranformation will be the object of the next section.

\section{Maxwell's multipoles}

In this section, Maxwell's multipole construction (\ref{eq:maxmulti1}) for a real function $f(\theta, \phi)$ of spin $j$ will be proved based on the decompositions (\ref{eq:kerdecomp}), (\ref{eq:pfdecomp}).

$f(\theta, \phi)$ may be obtained from integrating $p_f(\zeta)$ over the $\zeta$-plane using the inverse kernel $K^{\dagger}(\theta, \phi|\zeta),$ i.e. 
\begin{eqnarray}
   f(\theta,\phi) & = \int_{\mathbb{C}} \frac{\rmd\zeta\rmd\zeta^{\ast}}{(1+|\zeta|^2)^2} p_f(\zeta) K^{\dagger}(\theta, \phi|\zeta) \nonumber \\
   & = \int_{\mathbb{C}} \frac{\rmd\zeta\rmd\zeta^{\ast}}{(1+|\zeta|^2)^2} \prod_{n=1}^j (\bi{u}_n \cdot \bnu(\zeta)) (\bnu(\zeta)^{\ast} \cdot \bi{u}(\theta,\phi)),
   \label{eq:sylv1}
\end{eqnarray}
where the inverse kernel has been decomposed as the conjugate of (\ref{eq:kerdecomp}).

This integral, over all stereographic directions $\zeta,$ depends on the nilpotent vectors $\bnu(\zeta).$
The $\zeta$ itself is now redundant, and the integral can be taken over all $\bnu$ vectors with nilpotence and normalisation conditions provided by $\delta$-functions, written as an integral in a form similar to usual coherent state integrals,
\begin{equation}
\fl   f(\theta, \phi) = C \int \rmd^3 \, \bnu\rmd^3 \bnu^{\ast} \, \delta(|\bnu|^2 - 1) \delta(\bnu \cdot \bnu) \delta(\bnu^{\ast} \cdot \bnu^{\ast})\prod_{n=1}^j  (\bi{u}_n \cdot \bnu) ( \bnu^{\ast} \cdot \bi{u}(\theta,\phi)).
   \label{eq:sylv2}
\end{equation}
The pair of $\delta$-functions with conjugate arguments are equivalent to pair of $\delta$-functions with real independent arguments.

Now, the overall magnitude of $\bnu$ does not affect the $\theta, \phi$ dependence on this integral, and the normalisation condition may be relaxed.
It is most convenient to replace this with a gaussian distribution for the absolute value $|\bnu|,$ giving
\begin{equation}
\fl   f(\theta,\phi) = C \int \rmd^3 \bnu\, \rmd^3 \bnu^{\ast} \, \exp(-|\bnu|^2/2) \delta(\bnu \cdot \bnu) \delta(\bnu^{\ast} \cdot \bnu^{\ast}) \prod_{n=1}^j  (\bi{u}_n \cdot \bnu) (\bnu^{\ast} \cdot \bi{u}(\theta,\phi)).
   \label{eq:sylvgauss}
\end{equation}
This gaussian integral will now be evaluated using standard techniques.

The pair of $\delta$-functions forcing nilpotence may be replaced by their Fourier representation, in the (complex) Fourier variable $\kappa.$
Writing the total integral compactly, with $\bi{V} = (\bnu, \bnu^{\ast}),$
\begin{equation}
\fl   f(\theta,\phi) = C \int \rmd \kappa \, \rmd \kappa^{\ast} \int \rmd^3 \bnu\,\rmd^3\bnu^{\ast}  \exp(-\bi{V}^{\ast} \cdot \mathbf{M}\cdot\bi{V}) \prod_{n=1}^j   (\bi{u}_n \cdot \bnu) (\bnu^{\ast} \cdot \bi{u}(\theta,\phi)),
   \label{eq:sylvgauss1}
\end{equation}
where the $6\times 6$ matrix $\mathbf{M}$ in the exponent is given by
\begin{equation}
   \mathbf{M} = \frac{1}{4} \left( \begin{array}{cccccc} 1  & 0 & 0 & \rmi \kappa & 0 & 0 \\ 0 & 1 & 0 & 0 & \rmi \kappa & 0\\ 0 & 0 & 1 & 0 & 0 & \rmi \kappa \\ \rmi \kappa^{\ast} & 0 & 0 & 1 & 0 & 0 \\ 0 & \rmi \kappa^{\ast} & 0 & 0 & 1 & 0 \\ 0 & 0 & \rmi \kappa^{\ast} & 0 & 0 & 1 \end{array}\right).
   \label{eq:matrix}
\end{equation}
The inverse $\mathbf{M}^{-1}$ is the same as $\mathbf{M}$ with $\rmi$ replaced by $-\rmi$ and divided through by $(1+|\kappa|^2).$
It is now possible to find the gaussian integral over $\bnu, \bnu^{\ast}.$

This integral is possible by gaussian integration by parts (`Wick's theorem' in quantum field theory).
Up to a multiplicative constant, the gaussian integral is $(\det \mathbf{M})^{-1/2}=(1+|\kappa|^2)^{-3/2}$ times a combinatorial term, which is now described.

This combinatorial term is a sum over all possible pairings involving $\bnu$ and $\bnu^{\ast}$ in the product in (\ref{eq:sylvgauss1}).
Each summand is the product of the coefficients from the pairing (in terms of the inner product of the vectors $\bi{u}_n$ and $\bi{u}(\theta,\phi)$) times the component of $\mathbf{M}^{-1}$ relating the relevant pairing: $1/(1+|\kappa|^2)$ for $\bnu$ with $\bnu^{\ast},$ $-\rmi \kappa^{\ast}/(1+|\kappa|^2)$ for $\bnu$ with $\bnu,$ and $-\rmi \kappa/(1+|\kappa|^2)$ for $\bnu^{\ast}$ with $\bnu^{\ast}.$
The sum is therefore over the product of the dot product of the paired vectors, times the appropriate element of $\mathbf{M}^{-1}:$ i.e. $\bi{u}_n\cdot \bi{u}(\theta,\phi)/(1+|\kappa|^2),$ $\rmi \kappa \bi{u}(\theta,\phi)\cdot\bi{u}(\theta,\phi)/(1+|\kappa|^2),$ or $-\rmi \kappa^{\ast} \bi{u}_n\cdot \bi{u}_n.$

Clearly, there is only one distinct summand which pairs all the $\bi{u}(\theta,\phi)$ vectors with the $\bi{u}_n$ vectors; there is no $\kappa$ or $\kappa^{\ast}$ in the numerator of this term.
All of the other summands have at least one pairing of $\bi{u}(\theta,\phi)$ with itself, and two Majorana vectors paired.
With this in mind, (\ref{eq:sylvgauss}) becomes
\begin{equation}
   \fl f(\theta,\phi) = 
   C \left(\prod_{n=1}^j \bi{u}(\theta,\phi)\cdot\bi{u}_n \right) 
   \int \frac{\rmd \kappa \rmd \kappa^{\ast}}{(1+|\kappa|^2)^{j+3/2}} 
   + \bi{u}(\theta,\phi) \cdot \bi{u}(\theta,\phi) 
   \int\rmd\kappa \rmd \kappa^{\ast} F(\kappa,\kappa^\ast).
   \label{eq:sylv3}
\end{equation}
$F(\kappa,\kappa^{\ast})$ is the sum of all the remaining pairings.
The $\kappa, \kappa^{\ast}$ integrals may now be performed; the first summand gives a number, the second, a complicated (but unique) function involving $\bi{u}(\theta, \phi)$ and the $\bi{u}_n.$
This final integral therefore gives (\ref{eq:maxmulti1}), and the Maxwell multipoles have been demonstrated to be exactly the Majorana vectors of the function.

\section{Discussion and comparison with alternative proofs}\label{sec:comparison}

The main result of this paper is a proof of Sylvester's theorem (\ref{eq:maxmulti1}) using Majorana's sphere and the algebra of quantum spins. 
Having started with the real function $f(\theta, \phi),$ the strategy was to transform from the fully folded representation (basis functions of SO(3)) to the Majorana function (coherent basis functions of SL(2,$\mathbb{C}$)) in the fully extended representation, using the extending kernel. 
Using reality, the Majorana factorization is in terms of spin 1 states, rather than the more usual spin 1/2 states; for clarity, these were written in vector notation with the spin $+1$ coherent states represented by nilpotent vectors.
The transformation back to the folded representation was effected by transforming by the inverse kernel, which, as a real state, was also factorised.
The direction dependence of the final transformation was finally determined by transforming the integral on the sphere to a gaussian integral.
An additional feature of the derivation was the representation of the spherical harmonics via symmetric polynomials, between the first and second lines of (\ref{eq:kerdecomp}).

The proof provided by Backus \cite{backus:geometrical} is, in fact, rather similar to the one provided here.
He constructed the ring homomorphism from the space of homogeneous polynomials of order $j$ in three real variables $x, y, z,$ to the space of homogeneous polynomials of order $2j$ in two complex variables $\xi, \eta;$ the homomorphism he used (\cite{backus:geometrical} equation (50)) is precisely that obtained by using, as cartesian components, quadratic forms in $\xi$ and $\eta$ with respect to the the Pauli spin matrices times a real antisymmetric matrix. 
$\xi$ and $\eta$ are therefore the components of a spin 1/2 spinor, and the mapping is equivalent from going between the fully folded and fully extended representations.
His factorization of the complex polynomial, on replacing $\xi \to \zeta$ and $\eta \to 1$ is analogous to Majorana's theorem; reality of the original polynomial gives antipodality of the roots.
The transformation back is made using the first isomorphism theorem for (homogeneous polynomial) rings, by mapping into the quotient ring, factoring the original ring in $x,y,z$ with the kernel of the spinor mapping, i.e. homogeneous polynomials of the form $\bi{r} \cdot \bi{r} F.$
In the present proof, the transformation is realised concretely by integrating with respect to the inverse of the extending kernel.
Thus Backus's proof, using spinors in disguise, is equivalent to the present one.

The proof of Zou and Zheng \cite{zz:maxwell} is also similar, but in the language of tensors as used in continuum mechanics. 
Specifically, Sylvester's theorem is applied to traceless symmetric tensors (i.e. the unique irreducible tensor operators of maximal order $j$ in the decomposition of arbitrary tensors of rank $j$).
As with Backus's and the present proof, a complex polynomial is constructed by contracting with a complex tensor function of the complex variable $\zeta,$ whose complex roots occur in antipodal pairs.
Contracting with this tensor function is equivalent to transforming to the fully extended representation.
A further proof using the same general technique, but the different physical language of the spacetime spinor calculus described by Penrose and Rindler \cite{pr:spinors1, pr:spinors2}, is described in \ref{sec:spacetime}.
In this case, the objects satisfying Sylvester's theorem are symmetric hermitian spinors in 3 spatial dimensions (rather than 4 in spacetime).

Both Courant and Hilbert \cite{ch:methods1}, and Sylvester \cite{sylvester:note} use identities from the theory of algebraic curves, by directly complexifying $x, y, z$ space, and looking for the coincidences of the cones $f(x,y,z) = 0$ with $x^2 +y^2 + z^2 = 0.$ 
Here, Bezout's theorem is used to provide the existence of the multipole directions, rather than the fundamental theorem of algebra.

Majorana's sphere construction remains comparatively obscure in the quantum mechanics literature, despite providing a natural framework for describing the rotation of spins (and, in general, of functions on the sphere). 
It has been used here, through quantum mechanical language, to demonstrate a fundamental property of real states as multipoles, which provides a further role for the antipodal pairs of Majorana vectors.
It is striking to note that this philosophy and property of spinors goes back at least as far as Maxwell.

\section*{Acknowledgements} I am grateful to Gerald Kaiser, Iwo Bialynicki-Birula and Jonathan Robbins for discussions, to James Vickers for reading through \ref{sec:spacetime}, and particularly to John Hannay for much insight, and for introducing me to the Majorana sphere. This work was supported by the Leverhulme Trust.

\appendix

\section{The equivalence of the two Maxwell multipole representations (\ref{eq:maxmulti}), (\ref{eq:maxmulti1})}\label{sec:equiv}

It suffices to prove equivalence for a single term, that is,
\begin{equation}
   \partial_x^p \partial_y^q \partial_z^s \frac{1}{r} = C \frac{x^p y^q z^s}{r^{2j+1}}  + \frac{F}{r^{2j-1}}
   \label{eq:maxmult1}
\end{equation}
for $p+q+s = j,$ $C$ some constant independent of $p, q$ and $s,$ and $F$ some other homogeneous polynomial in $x,y,z$ of order $j-2.$

To begin, $r$ is replaced by its Fourier integral representation,
\begin{eqnarray}
   \partial_x^p \, \partial_y^q \, \partial_z^s 1/r 
   & = \partial_x^p \, \partial_y^q \, \partial_z^s \frac{1}{2\pi^2}\int \frac{\rmd^3\bi{k}}{k^2} \exp(\rmi \bi{k}\cdot \bi{r}) \nonumber \\
   & = \frac{\rmi^j}{2\pi^2} \int \frac{\rmd^3\bi{k}}{k^2} k_x^p k_y^q k_z^s \exp(\rmi \bi{k}\cdot \bi{r}) \nonumber \\
   & = \frac{\rmi^j}{4\pi^2} \int_0^{\infty}\rmd \lambda \int \rmd^3 \bi{k} \, k_x^p k_y^q k_z^s \exp(-k^2 \lambda/2  + \rmi \bi{k}\cdot \bi{r})
   \label{eq:maxmulti2}
\end{eqnarray}
where in the second line, the partial derivatives have been performed, and in the third, $1/k^2$ itself is represented as an integral  over $\lambda.$
The integrals in $k_x, k_y$ and $k_z$ can now be evaluated separately; following the appropriate linear transformation $k_x = t/\sqrt{\lambda} +\rmi x/\lambda,$ the $k_x$ integral transforms to
\begin{equation}
\fl   \int \rmd k_x \, k_x^p \exp(-k_x^2 \lambda/2 + \rmi k_x x) = \frac{\exp(-x^2/2\lambda)}{\lambda^{p+1/2}} \int \rmd t \, (\sqrt{\lambda} t + \rmi x)^p \exp(-t^2/2).
   \label{eq:maxmulti3}
\end{equation}
The bracketed term to the $p$th power in the integrand must now be expanded; all terms involving an odd power of $t$ vanish by symmetry, and the nonvanishing terms are real due to the $\rmi^j$ factor in the front of (\ref{eq:maxmulti2}).

The $(\rmi x)^p$ summand term may be integrated without problem as a usual gaussian (it is independent of $p$).
Multiplying this by the analogous $(\rmi y)^q$ and $(\rmi z)^s$ terms, this summand of (\ref{eq:maxmulti2}) gives, up to a constant factor
\begin{eqnarray}
   x^p y^q z^s \int \frac{\rmd\lambda}{\lambda^{j+3/2}} \exp(-r^2/2/\lambda) 
   & = \frac{x^p y^q z^s}{r^{2j+1}} \int \rmd \rho \, \rho^{j-1/2} \exp(-\rho/2) \nonumber \\ 
   & = C x^p y^q z^s /r^{2j+1}.
   \label{eq:maxmultis1}
\end{eqnarray}
In the first line, $\lambda$ has been replaced by $\rho = r^2/\lambda.$ 
This proves the first part of (\ref{eq:maxmulti1}).

In all of the summands not included in the expression (\ref{eq:maxmultis1}), the exponent of $\lambda$ in the denominator is less than $j+3/2$ by some integer (since only even powers of $t$ in (\ref{eq:maxmulti3}) contribute).
On transforming to $\rho,$ the power of $r$ in the resulting denominator is less by at least two; $1/r^{2j-1}$ therefore factors out, giving the form for the second summand in (\ref{eq:maxmulti1}).
Therefore, the two forms of Maxwell's multipole expansion are the same.

\section{Sylvester's theorem using spacetime spinor calculus}\label{sec:spacetime}

A further proof of Sylvester's theorem, similar in form to that proved in the main text, and \cite{backus:geometrical, zz:maxwell}, will be outlined here using the calculus of relativistic spinors, as described by Penrose and Rindler \cite{pr:spinors1, pr:spinors2}, referred to as I and II in this appendix.
It should be noted that if algebra of spinors in 3-space were to be used from the outset, as in \cite{ll:quantum} chapter VIII, leads to a more straightforward demonstration (see, in particular, the final section of this chapter).

Using index notation and employing the summation convention, the function $f(\theta, \phi)$ with total angular momentum number $j$ may be written
\begin{equation}
   f(\theta, \phi) = T_{a \cdots g} u^a(\theta, \phi) \cdots u^g(\theta, \phi),
   \label{eq:tensequiv}
\end{equation}
where $u^a(\theta, \phi)$ denote the components of the unit vector $\bi{u}(\theta,\phi)$ (and, in the following, the vector itself).
In (\ref{eq:tensequiv}), $T_{a \cdots g} = T_{(a \cdots g)}$ denotes a real, completely symmetric rank $j$ tensor (with round brackets $(\cdots )$ denoting complete symmetrization, as usual), which is traceless with respect to contraction with any pair of indices.
Where convenient, it will be assumed that the basis is fixed.

It is important to note that to use the spacetime spinor calculus, the indices in (\ref{eq:tensequiv}) have to be over 4-dimensional minkowskian spacetime (with signature $-2$).
However, Maxwell's multipoles are a feature of 3-dimensional space only, so it is assumed that there is an absolute newtonian time direction, that is, a timelike 4-vector $v^a = (\sqrt{2},0,0,0),$ and $T_{a \cdots g}, u^a(\theta, \phi)$ lie on the 3-dimensional spacelike hypersurface orthogonal to $v^a.$

All tensor indices $a, b, \dots$ can be replaced by pairs of spinor indices $A A', B B', \dots,$ using the Infeld-van der Waerden symbols $g_{a}^{\;AA'};$ in the present case, these are simply the three Pauli matrices and the identity, with a $2^{-1/2}$ prefactor (I section 3.1).
The primed indices are complex conjugated coordinates, and spinor indices are raised and lowered using the antisymmetric symbols $\varepsilon_{AB}, \varepsilon^{AB}, \varepsilon_{A'B'}, \varepsilon^{A'B'},$ where $\varepsilon_{01} = \varepsilon_{0'1'} =  \varepsilon^{01} =  \varepsilon^{0'1'} = +1$ (I section 2.5).

Thus, for instance,
\begin{equation}
   v_a v^a = v_{A A'} v^{A A'} = 2,
   \label{eq:vnorm}
\end{equation}
implying that $v^{AA'}$ is simply the identity matrix $\sqrt{2} g_0^{\;AA'}.$
Therefore (II p461), $v_A^{A'}, v^A_{A'}$ may be used to convert between primed and unprimed indices; the relevant components are
\begin{equation}
   v_0^{0'} = v^0_{0'} = v_1^{1'} = v^1_{1'} = 0, \quad v_1^{0'} = v^1_{0'} = +1, \quad v_0^{1'} = v^0_{1'} = -1.
   \label{rlcomp}
\end{equation}

Since the tensor $T_{a \cdots g}$ is symmetric and traceless, the corresponding spinor $T_{A \cdots G A' \cdots G'} = T_{(A \cdots G) (A' \cdots G')}$ is totally symmetric in its primed and its unprimed indices (I p146).
Furthermore, since the tensor is real, the spinor is hermitian:
\begin{equation}
   \left[ T_{(A\cdots G)(A' \cdots G')} \right]^{\ast}
   = T^{\ast}_{(A'\cdots G')(A \cdots G)}
   = T^{\ast}_{(A\cdots G)(A' \cdots G')}
   = T_{(A\cdots G)(A' \cdots G')}.
   \label{eq:therm}
\end{equation}
In the second line, the indices have been rearranged (since the ordering of primed indices with respect to unprimed is unimportant), and in the third, reality of the components of $T$ was used (I pp123-4).
Since $T$ is orthogonal to the timelike vector $v,$ in terms of spinors,
\begin{equation}
   T_{(A\cdots G)(A'\cdots G')}v^{A A'} = 0.
   \label{eq:Tspace}
\end{equation}

This spinor may be expressed completely in unprimed indices,
\begin{equation}
   \tau_{(A\cdots G)(A_1 \cdots G_1)} = T_{(A\cdots G)(A' \cdots G')} v^{A'}_{A_1} \cdots v^{G'}_{G_1}.
   \label{eq:taudef}
\end{equation}
Contracting any two indices from each of the two sets gives
\begin{eqnarray}
   \tau_{(A\cdots G)(A_1\cdots G_1)} \varepsilon^{AA_1} 
   & = T_{(A\cdots G)(A'\cdots G')} v^{A'}_{A_1} \varepsilon^{AA_1} v^{B'}_{B_1} \cdots v^{G'}_{G_1} \nonumber \\
   & = -T_{(A\cdots G)(A'\cdots G')} v^{A A'} v^{B'}_{B_1} \cdots v^{G'}_{G_1} \nonumber \\
   & = 0
   \label{eq:tausym}
\end{eqnarray}
by (\ref{eq:Tspace}).
This shows that $\tau$ is symmetric with respect to indices in its two symmetrised sets, and thus is totally symmetric: $\tau_{(A\cdots G)(A_1\cdots G_1)}=\tau_{(A\cdots G A_1\cdots G_1)}.$

This totally symmetric spinor may be canonically decomposed (I proposition (3.5.18)),
\begin{equation}
   \tau_{(A \cdots G A_1 \cdots G_1)} = \alpha_{(A} \cdots \gamma_{G} \alpha'_{A_1} \cdots \gamma'_{G_1)},
   \label{eq:candec}
\end{equation}
where spinors are primed when their index has subscript 1.
This step is, of course, equivalent to Majorana's theorem, which is proved (I p162) by using the fundamental theorem of algebra on the order $2j$ (Majorana) polynomial
\begin{equation}
   \tau(\zeta) = \tau_{(A \cdots G_1)} \xi^{A} \cdots \xi^{G_1} \quad \mathrm{where} \quad \xi^A = \left( \begin{array}{c} 1 \\ \zeta \end{array} \right).
   \label{eq:spinorpoly}
\end{equation}
and a spin frame is chosen where the complex number $\zeta \ne 0.$

It is necessary to show that the roots of this polynomial are antipodal, since then the spinors in the decomposition (\ref{eq:candec}) are arranged in antipodal pairs, which will be demonstrated similarly to (\ref{eq:repol}).
The spinors $\xi^A$ may be replaced by conjugated, primed spinors $\eta^{\ast A'}$ using $v_A^{A'},$
\begin{equation}
   \eta^{\ast A'} = -v^{A'}_A \xi^A.
   \label{eq:etadef}
\end{equation}
From $\xi^A$ defined in (\ref{eq:spinorpoly}), (\ref{eq:etadef}) gives
\begin{equation}
   \eta^A = \left(\begin{array}{c} -\zeta^{\ast} \\ 1 \end{array}\right) \propto \left(\begin{array}{c} 1 \\ -1/\zeta^{\ast} \end{array}\right).
   \label{eq:etadef2}
\end{equation}
The polynomial (\ref{eq:spinorpoly}) may thus be written in terms of $\xi$ and $\eta,$
\begin{equation}
   \tau(\zeta) = (-1)^j T_{(A\cdots G)(A'\cdots G')} \xi^A \cdots \xi^G \eta^{\ast A'} \cdots \eta^{\ast G'}.
   \label{eq:spinorpoly1}
\end{equation}
Since the components of $T$ are real, conjugation of the polynomial gives the same result as in (\ref{eq:repol}):
\begin{eqnarray}
   \tau(\zeta)^{\ast} & = (-1)^j T^{\ast}_{(A'\cdots G')(A\cdots G)} \xi^{\ast A'} \cdots \xi^{\ast G'} \eta^A \cdots \eta^G \nonumber \\
   & = (-1)^j T_{(A\cdots G)(A'\cdots G')}  \eta^A \cdots \eta^G \xi^{\ast A'} \cdots \xi^{\ast G'} \nonumber \\
   & \propto \tau(-1/\zeta^{\ast}).
   \label{eq:repol1}
\end{eqnarray}
Thus the roots are antipodal.

Transforming the Maxwell multipole form of (\ref{eq:maxmulti1}) into spinor form is equivalent to $\tau_{(A \cdots G_1)},$ in the sense of I p140: it equals it, plus combinations of lower rank spinors combined with the antisymmetric symbol.
This equivalence relation is equivalent to ignoring the second summand in the multipole decomposition (\ref{eq:maxmulti1}).
Thus Sylvester's theorem is proved.


\end{document}